\documentclass[conference]{IEEEtran}
\IEEEoverridecommandlockouts
\usepackage{cite}
\usepackage{amsmath,amssymb,amsfonts}
\usepackage{algorithmic}
\usepackage{graphicx}
\usepackage{textcomp}
\usepackage{xcolor}

\def\BibTeX{{\rm B\kern-.05em{\sc i\kern-.025em b}\kern-.08em
    T\kern-.1667em\lower.7ex\hbox{E}\kern-.125emX}}

\begin{document}

\title{Enhancing Thrust in Flapping Airfoils Through Wake Interactions with Oscillating Cylinder}

\author{\IEEEauthorblockN{Amir Khan}
\IEEEauthorblockA{\textit{Department of Mechanical Engineering} \\
\textit{NUST, CEME}\\
Rawalpindi, Pakistan \\
akhan.me20ceme@student.nust.edu.pk}
\and
\IEEEauthorblockN{Imran Akhtar}
\IEEEauthorblockA{\textit{Department of Mechanical Engineering} \\
\textit{NUST, CEME}\\
Rawalpindi, Pakistan \\
imran.akhtar@ceme.nust.edu.pk}
\and
\IEEEauthorblockN{Muhammad Saif Ullah Khalid}
\IEEEauthorblockA{\textit{Department of Mechanical Engineering} \\
\textit{Lakehead University}\\
Thunder Bay, ON, Canada \\
mkhalid7@lakeheadu.ca}
}

\maketitle

\begin{abstract}
Inspired by the natural motion of insects, fish, and other animals, flapping airfoils have gained significant importance due to their applications in fields such as ship propulsion, micro aerial vehicles, and autonomous underwater vehicles. Over the past two decades, extensive research has focused on understanding the dynamics of these airfoils, their thrust production capabilities, and methods to enhance this thrust in unsteady flows. This study investigates how the presence of a cylinder oscillating due to incoming flow affects the thrust performance of the flapping airfoil. The results indicate that the flapping airfoil generates increased thrust when placed in the wake of the oscillating cylinder compared to a scenario without the cylinder's wake.  A direct relationship has been found between the strouhal number ($St$) of the flapping airfoil and its pitching amplitude, which significantly influences the airfoil's performance. This analysis highlights the potential for optimizing flapping airfoil efficiency through strategic selection of flapping parameters and the placement of an oscillating cylinder.
\end{abstract}


\section{Introduction}

Salmon, tuna, sharks, dragonflies, hawks, moths, and other similar animals use oscillation to propel themselves. These animals use fins or wings to overcome cruising drag. Understanding oscillatory propulsion's physical mechanisms is necessary for high-performance AUVs and MAVs. The oscillating foil is a simplified model for oscillatory propulsion, and flapping motion thrust generation has been extensively studied. Review articles on this topic are available from Wu \cite{wu2011fish} and Triantafyllou et al. \cite{triantafyllou_techet_hover_2004}.\\

Flapping foil plunge and pitch oscillate in the oscillating-airfoil model. Research indicates that combining plunge and pitch oscillations outperforms a single DOF motion \cite{ramamurti2001simulation}. Many studies show that a 90-degree plunge-pitch phase difference maximizes efficiency. Studies by Isogai et al. \cite{isogai1999effects}, Ramamurti and Sandberg \cite{ramamurti2001simulation}, Young et al. \cite{young2014review}, and Van Buren et al. \cite{van2019scaling} support this. Van Buren et al. \cite{van2019scaling} found that thrust is highest at phase angles below 90 degrees. Anderson et al. \cite{anderson1998oscillating} achieved 0.87 propulsive efficiency in their complex experiments on oscillating foils. Their flow visualization experiments showed that high efficiency required a moderately intense leading-edge vortex (LEV). A jet-like averaged flow and reverse von Kármán vortex street were observed behind the oscillating foil during thrust production \cite{triantafyllou1993optimal, gopalkrishnan1994active}. Lai and Platzer \cite{lai1999jet} carried out experiments on jet flows during the drag-to-thrust transition.\\

Strouhal number is a key factor in flapping foil propulsivity. An analysis by Streitlien and Triantafyllou \cite{streitlien1998thrust} indicates that thrust rises with Strouhal number or oscillation frequency. But high reduced frequencies can reduce efficiency. The ideal Strouhal number for fish swimming is between 0.25 and 0.4, as shown in natural and experimental studies \cite{anderson1998oscillating, triantafyllou1993optimal}. Advanced research on thrust generation in oscillating foils has led to practical applications in MAVs \cite{triantafyllou_techet_hover_2004}.\\

While the mechanisms of thrust generation in oscillating foils are well understood, they cannot fully explain the high propulsive performance of natural creatures, which can achieve both high thrust and efficiency simultaneously \cite{triantafyllou_techet_hover_2004}. Most plunging and pitching studies used 90 degrees phase angles. Anderson et al. \cite{anderson1998oscillating} measured a maximum thrust of 1.0 and an efficiency of 0.6, significantly lower than the reported 0.87 efficiency. In contrast, Van Buren et al. \cite{van2019scaling} reported a thrust value above 15.0 but a low efficiency of 0.2. Comparing Table I shows that increasing oscillation frequency or amplitude to increase thrust reduces propulsive efficiency. Compared to other studies, Isogai et al. \cite{isogai1999effects} found an efficiency of 0.75 with sufficient thrust, but with lower oscillation frequency and higher plunging amplitude The key challenge is to achieve high thrust while maintaining propulsive efficiency, which motivates this study to understand natural creatures' propulsive behaviors.\\

Most studies covered oscillating airfoils in uniform oncoming flows. However, practical scenarios often involve nonuniform oncoming flow conditions, like unsteady vortices \cite{triantafyllou_techet_hover_2004}. Fish schooling and fin-fin interactions can conserve energy through complex vortex/fin interactions \cite{weihs_1973}. Triantafyllou et al. \cite{triantafyllou_techet_hover_2004} discovered that body-generated vortices on the flapping caudal fin can enhance oscillatory swimming performance, highlighting the need for further research. Gopalkrishnan et al. (1994) used an oscillating foil to control the wake of a circular cylinder, identifying three typical wake modes through flow visualization. Foil wake change suggested fluid forces may vary. Interestingly, cylinder wake vortices can propel a dead fish upstream \cite{triantafyllou1993optimal}. In a study by Liao et al. \cite{liao2003fish}, fish moving through vortices reduced muscle activity and energy cost for thrust generation. Experiments show that oncoming vortices significantly affect thrust generation, fluid forces, and energy exchange in oscillating-foil propellers.\\

According to Yuan and Hu \cite{yuan2017numerical}, strong interactions between oncoming vortices and oscillating foil can increase thrust and energy consumption. Guo et al. \cite{guo_han_zhang_wang_lauder_valentina_di_santo_dong_2023} used particle imaging velocimetry (PIV) to study the interaction between vortices from an upstream anal fin and a downstream caudal fin. They found that upstream vortices' LEV stabilized and reinforced caudal fin LEV formation, a thrust enhancement mechanism. Poudel et al. \cite{poudel2023impacts} found that increasing the unsteadiness of oncoming flow with an upstream array of cylinders resulted in more thrust enhancement cases than reduced thrust cases in simulations\\

Poudel et al. found a constructive vortex pattern and thrust enhancement. \cite{poudel2023impacts} and Gopalkrishnan et al. \cite{gopalkrishnan1994active} recommended both classic and reverse von Kármán streets under similar conditions. This discrepancy may be due to oscillation parameters and flow conditions, highlighting the need for thrust enhancement wake pattern research. \\

This study examines performance of flapping foil in terms of thrust and lift/side-forces from incoming vortices. Vortices are generated by Vortex-induced oscillations of a circular cylinder upstream from the airfoil, previous studies were conducted in the vortices of a stationary bluff body \cite{liao2004vortex, shao2011hydrodynamics, shao2010hydrodynamic, yuan2017numerical, li2021investigation}. The structural parameters of the oscillating cylinder are kept constant throughout the study to only focus on the performance of the airfoil.\\

\section{Methodology}
In this study, numerical simulations were conducted using the two-dimensional, incompressible, and unsteady Navier-Stokes equations in Cartesian coordinates with ANSYS Fluent 2023 R1. The simulations employed the continuity equation and momentum equations.

\begin{equation} \frac{\partial \bar{u_i}}{\partial x_i} = 0 \label{eq:continuity} \end{equation}

\begin{equation} \frac{\partial \rho \bar{u_i}}{\partial t} + \frac{\partial \rho \bar{u_i} \bar{u_j}}{\partial x_i} = -\frac{\partial \bar{p}}{\partial x_i} + \mu \nabla^2 \bar{u_i} - \frac{\partial \rho u'_i u_j}{\partial x_j} \label{eq:momentum} \end{equation}

The pressure-based coupled algorithm to solve the equations, ensuring stability and accuracy in the computation of the pressure-velocity coupling. To handle high mesh displacements, particularly with moving bodies such as flapping foils, the overset meshing technique was employed. This approach avoids the complications of mesh deformation in dynamic mesh methods by preserving the mesh shape and improving the accuracy of the simulation results.\\

For the fluid-structure interaction (FSI) involving the flapping foil and vortex-induced vibrations (VIV) of the cylinder, a user-defined function (UDF) was developed. This UDF facilitates the coupling between the fluid and structural solvers, allowing for accurate simulation of the dynamic interactions and kinematics of the flapping motion.\\

\subsection{Flapping Kinematics}

This work investigates a NACA0012 airfoil that exhibits oscillations in both plunge and pitch. The sinusoidal prescription of the plunge and pitch displacements, respectively indicated as $A^*$ and $h$, is given by Equations \ref{eq:H} and \ref{eq:TH}, where $\phi$ represents the phase difference. In order to improve propulsive efficiency, the angle $\phi$ is fixed to 90° in this work. The amplitude of oscillation in plunge, normalized with respect to the chord length c of the foil, is represented by $A^*_0$, whereas $h_0$ denotes the oscillation amplitude in pitch, measured in degrees. It is established that the pivot point is located precisely one-third of the chord length from the leading edge (LE) of the foil. The distance from the pivot point to the center of the cylinder is $2D$. Definition of the reduced frequency of the foil's oscillation, $f^*$, is given by Equation \ref{eq:FR}, where $U_{\infty}$ represents the inlet velocity. Normalized time is defined by Equation \ref{eq:T}. The present work assigns values of 0.5 and 10°, 20° and 30° to $A^*_0$ and $\theta_0$ correspondingly. The range of $f^*$ is defined as [0.1, 0.6] and $A^*_0$ is kept constant at 0.5, which also makes Strouhal number within the interval of [0.1, 0.6] as per Equation \ref{eq:ST}. Based on the research conducted by Kinsey et. al \cite{kinsey2008parametric}, a single foil in a uniform approaching flow undergoes a transition from a drag to a thrust condition as the $(f^*)$ grows within the defined frequency range, considering the oscillation amplitudes used.

\begin{equation}
    A^* = A_0^* \sin(2\pi f^* t^*)
    \label{eq:H}
\end{equation}
\vspace{-4mm}
\begin{equation}
    \theta = \theta_0 \sin(2\pi f^* t^* - \phi)
    \label{eq:TH}
\end{equation}
\vspace{-4mm}
\begin{equation}
    f^* = \frac{f c}{U_\infty}
    \label{eq:FR}
\end{equation}
\vspace{-3mm}
\begin{equation}
    t^* = \frac{t U_\infty}{c}
    \label{eq:T}
\end{equation}
\vspace{-3mm}
\begin{equation}
    St = \frac{2 f A_0^*}{U_\infty}
    \label{eq:ST}
\end{equation}

\subsection{Vortex Induced Vibrations of Circular Cylinder}
\begin{figure}[hb!]
    \centering
    \includegraphics[width=0.75\linewidth]{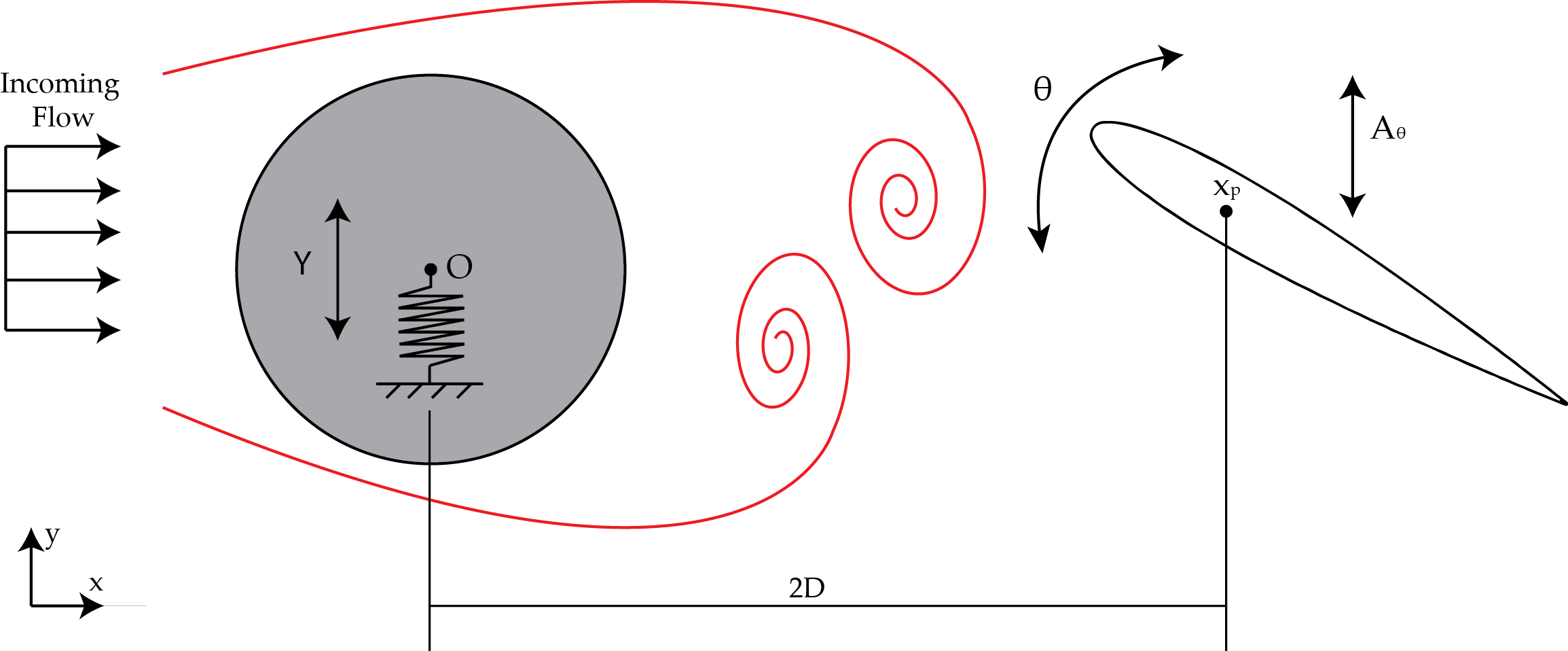}
    \caption{Illustration of an flapping NACA0012 airfoil positioned at a distance 2D in wake of a circular cylinder oscillating due to vortex-induced vibration.}
    \label{fig:illustration}
\end{figure}

Upstream vortices are generated by a laterally oscillating circular cylinder, governed by Equation \ref{eq:viv}. Traditional models include fluid-induced forces on the airfoil within this equation, omitting additional mass effects. Previous studies indicate that the average non-dimensional added mass remains consistent across different fluids and cylinder models \cite{chen_md_mahbub_alam_zhou_2020}. Figure \ref{fig:illustration} illustrates the interaction between the cylinder and the oscillating foil, with \(\theta\) defined as positive in the counterclockwise direction. In this study, the cylinder's diameter \(D\) is equal to the airfoil’s chord length \(c\). The foil's pivot x-coordinate and the cylinder's distance \(S\) are kept constant, while the cylinder's y-coordinate is \(A^* = 0.5\). The Reynolds number, based on the foil's longitudinal dimension, is set to 1000. Extensive research on oscillating foils at this Reynolds number \cite{kinsey2008parametric,ashraf_young_lai_2012,hoke_young_joseph_2023} provides a strong foundation for understanding how upstream vortices influence foil performance.

\begin{equation}
    \Ddot{Y}+2 \zeta_Y\left(\frac{2\pi}{U_r^*}\right)\dot{Y}+\left(\frac{2\pi}{U_r^*}\right)^2 Y=\frac{2}{\pi m^*}C_L
    \label{eq:viv}
\end{equation}

\( \zeta_y = c_y / c_{crit} = c_y / 2 \sqrt{k_y m_y} \) represents the damping ratio for the cylinder's vertical movement, while \( m^* = m_y / m_f \) indicates the ratio of the cylinder's mass per unit length \( m_y \) to the fluid mass \( m_f = \rho \pi L^2 / 4 \). For our study, the damping is kept as 0.

\subsection{Computational Domain}

\begin{figure}[hb!]
\centering
\includegraphics[width= 0.45 \textwidth]{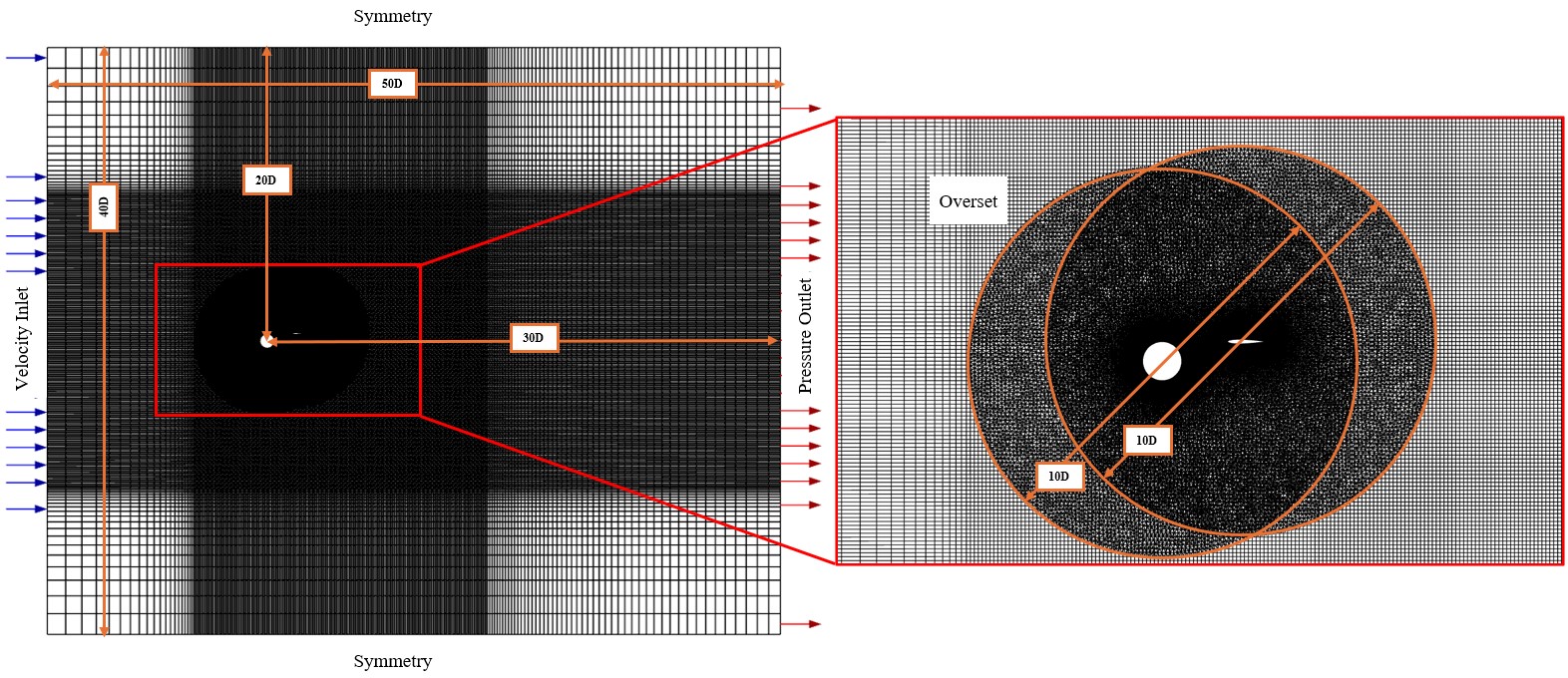}
\caption{Overset mesh with the structured background mesh and two unstructured meshes of cylinder and airfoil in front }
\label{fig:Domain1}
\end{figure}

A rectangular computational domain with dimensions 40D (height) by 50D (width), where D is the cylinder diameter, serves as the background mesh for this study. A velocity inlet is placed on the left, and a pressure outlet is positioned on the right to ensure fully developed flow. The top and bottom boundaries are symmetry conditions. The geometry of the region of interest, along with the boundary conditions, is illustrated in Figure \ref{fig:Domain1}.
\\

Superimposed on this background mesh are component meshes for a cylinder and an airfoil. The component mesh of cylinder is unstructured circular mesh, with a diameter of 10D, is centered at (0,0), while the airfoil, also with a diameter of 10D, is located at a distance 2D from the cylinder’s center, with its pivot one-third of the chord length from the leading edge.

\subsection{Flow Solver Validation}
\begin{figure}[hb!]
\centering
\includegraphics[width=0.35 \textwidth]{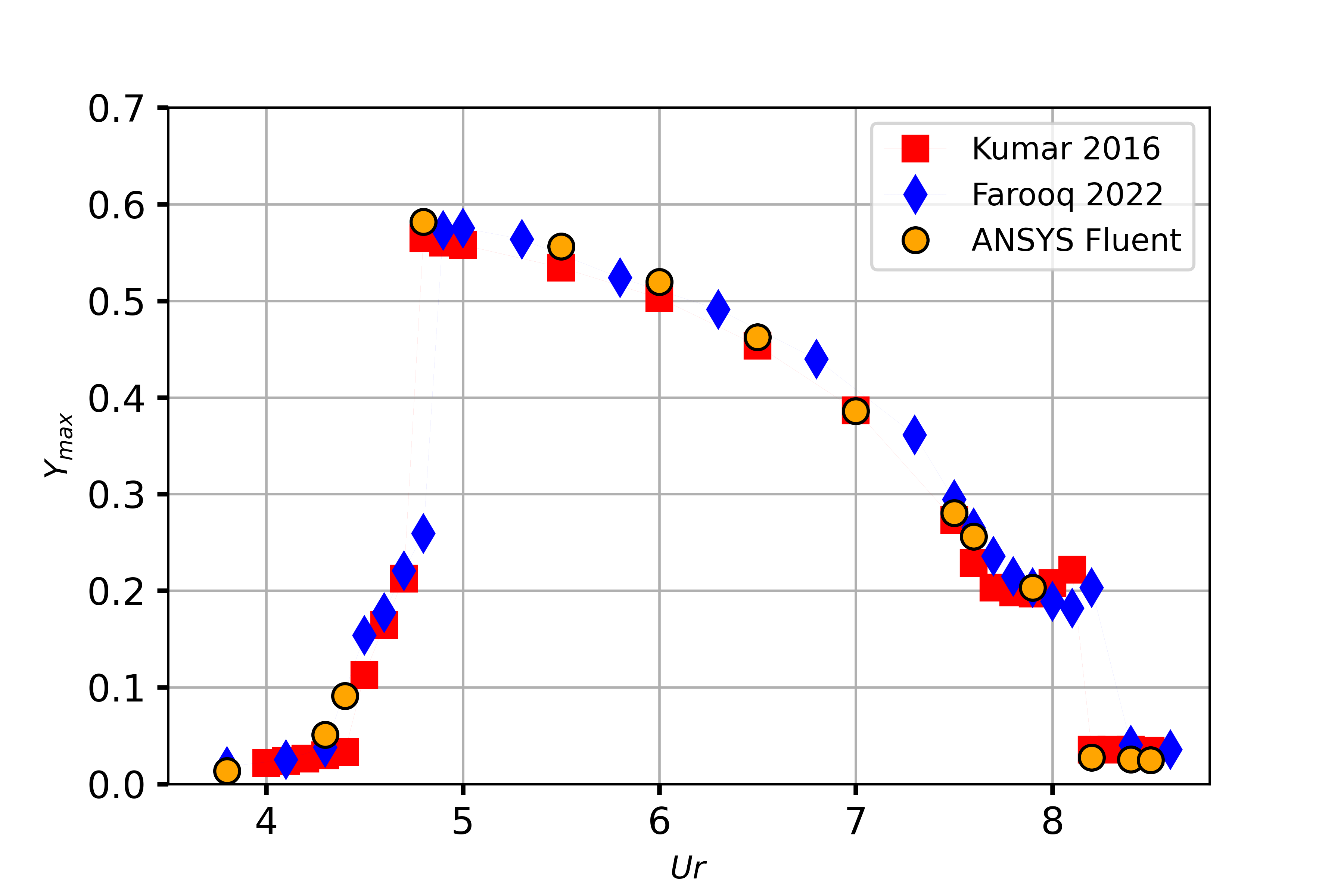}
\caption{Solver validation for VIV cylinder showing close correspondence with similar studies conducted \cite{kumar_navrose_mittal_2016,farooq_khalid_akhtar_hemmati_2022,farooq_muhammad_akhtar_hemmati_2023}}
\label{fig:Validation}
\end{figure}

To verify the accuracy and reliability of our solver, we performed a series of validation tests by reproducing the results from the research conducted by Kumar et al. \cite{kumar_navrose_mittal_2016} and Farooq et al. \cite{farooq_khalid_akhtar_hemmati_2022,farooq_muhammad_akhtar_hemmati_2023}. The experiments were conducted at a Reynolds number of 100 and a damping ratio ($\zeta$) of 0, with variations in the reduced velocity ($U_r^\ast$). We concentrated on showing the maximum vibration amplitude in the y-direction ($Y_{max}$) with the findings from these prior investigations as seen in Figure \ref{fig:Validation}. Our results shown a robust correlation with those documented by Kumar et al. \cite{kumar_navrose_mittal_2016} and Farooq et al. \cite{farooq_khalid_akhtar_hemmati_2022,farooq_muhammad_akhtar_hemmati_2023}, hence affirming the reliability and precision of our solver. This validation method confirms the solver's capacity to precisely simulate and analyze the system's dynamics under the stated conditions.\\

\begin{figure}[hb!]
\centering
\includegraphics[width=0.5 \textwidth]{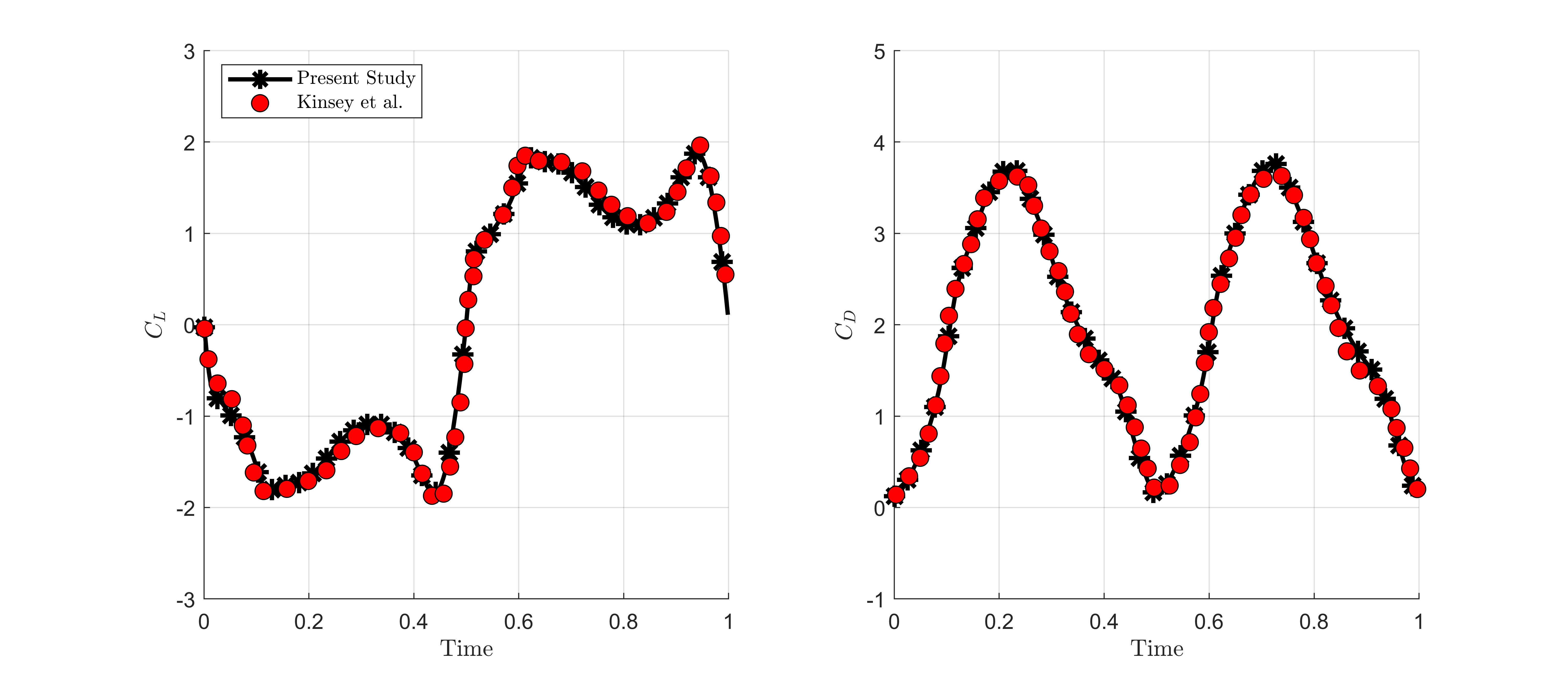}
\caption{Grid used in the present study showing strong correspondence with the study conducting by Kinsey et. al\cite{kinsey2008parametric}}
\label{fig:Kinsey}
\end{figure}

A study on grid independence and temporal independence was conducted utilizing the parameters established by Kinsey et al. \cite{kinsey2008parametric} The results illustrated in Figure \ref{fig:Kinsey} demonstrate that our mesh and time step sizes correspond effectively with those employed in Kinsey et al.'s research \cite{kinsey2008parametric}. This agreement indicates that the selected mesh and time step sizes are suitable for the simulation. The time step size was established at 2000 steps each frequency period, ensuring sufficient refinement of both mesh and temporal resolution for precise outcomes.\\

To summarize, the parameters included in this study are Reynolds number of 1000 and a Strouhal number range of 0.10 to 0.60. The heaving amplitude was set at 0.5, with pitching amplitudes of 10°, 20°, and 30°. The reduced velocity was maintained at 5.0, and the damping ratio was 0.0. The mass ratio was 2.0. Lift and thrust coefficients were calculated using the formulas \( C_L = F_y / (0.5 \rho_\infty U_\infty^2 c L) \) and \( C_T = -F_x / (0.5 \rho_\infty U_\infty^2 c L) \).

\section{Results \& Discussion}
\begin{figure}[hb!]
    \centering
    \includegraphics[width=1\linewidth]{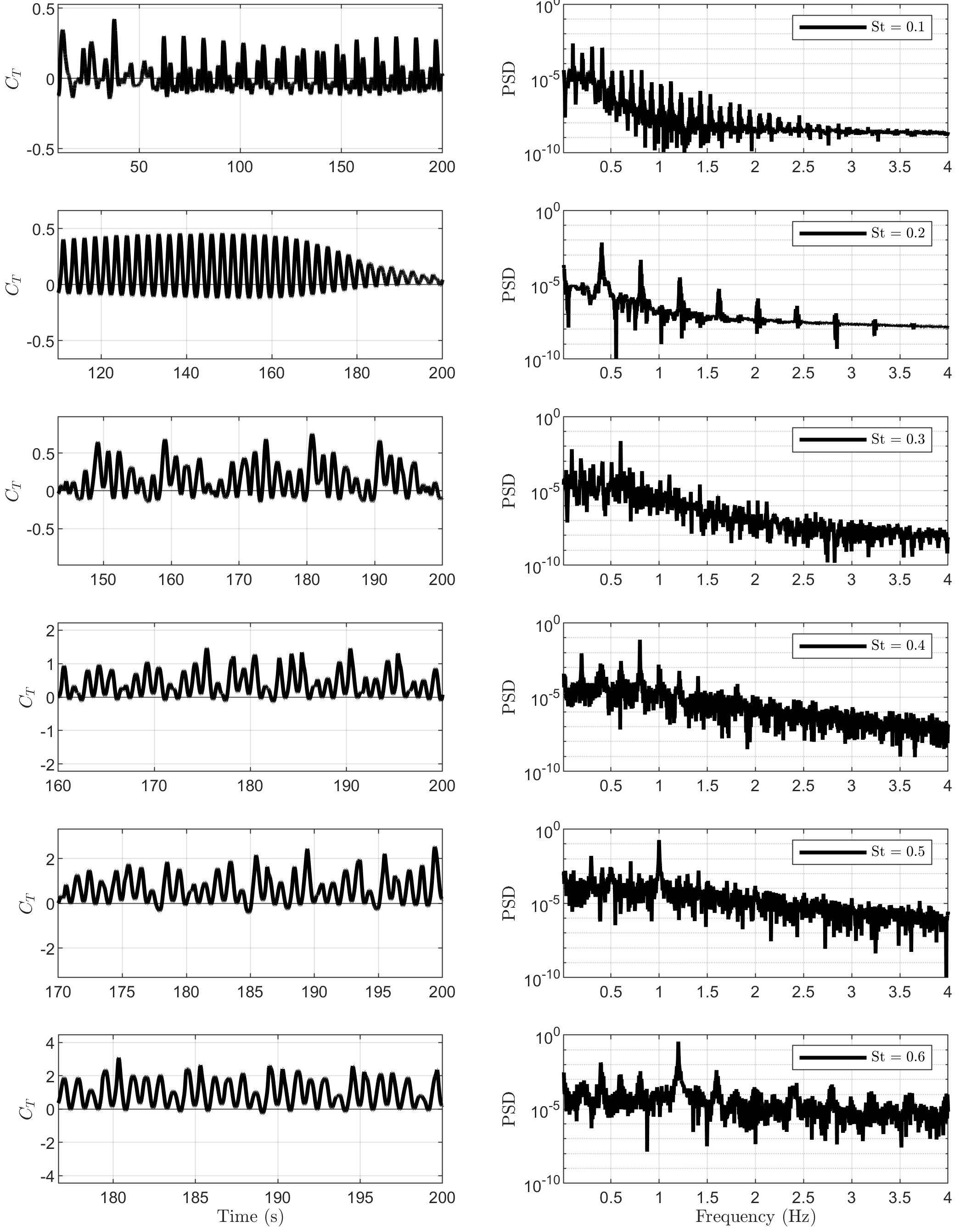}
    \caption{Time histories and FFT of $C_T$ for different Strouhal number. }
    \label{fig:CT_FFT}
\end{figure}

In this study, baseline values were established by performing flapping airfoil simulations with varying Strouhal numbers, ranging from 0.1 to 0.6 with different pitching amplitude of 10$^\circ$, 20$^\circ$ and 30$^\circ$. These simulations were conducted in the absence of upstream wakes or vortices. A cylinder, positioned at a distance of 2D from the airfoil as shown in Figure \ref{fig:illustration}, was subjected to oscillations induced by the incoming flow, generating vortex-induced vibrations. These vortices subsequently interacted with the flapping airfoil. Simulations were executed with the vortex-induced vibration (VIV) cylinder placed upstream of the flapping airfoil. The results from these simulations are discussed in the following section. Figure \ref{fig:CT_FFT} shows the time and frequency response of flapping at different Strouhal numbers.\\

The root mean square (RMS) was employed rather than the mean to compute the average, so providing a more precise representation of the oscillatory forces in the study. The mean was considered inadequate because it fails to represent oscillatory behavior. RMS is preferred as it considers both positive and negative oscillations, yielding a more precise assessment of amplitude. In oscillatory systems, the mean may converge to zero if positive and negative fluctuations negate each other, resulting in a distortion of system behavior.

\subsection{Performance Dependence on Pitching Amplitude}
\begin{figure}
    \centering
    \includegraphics[width=1\linewidth]{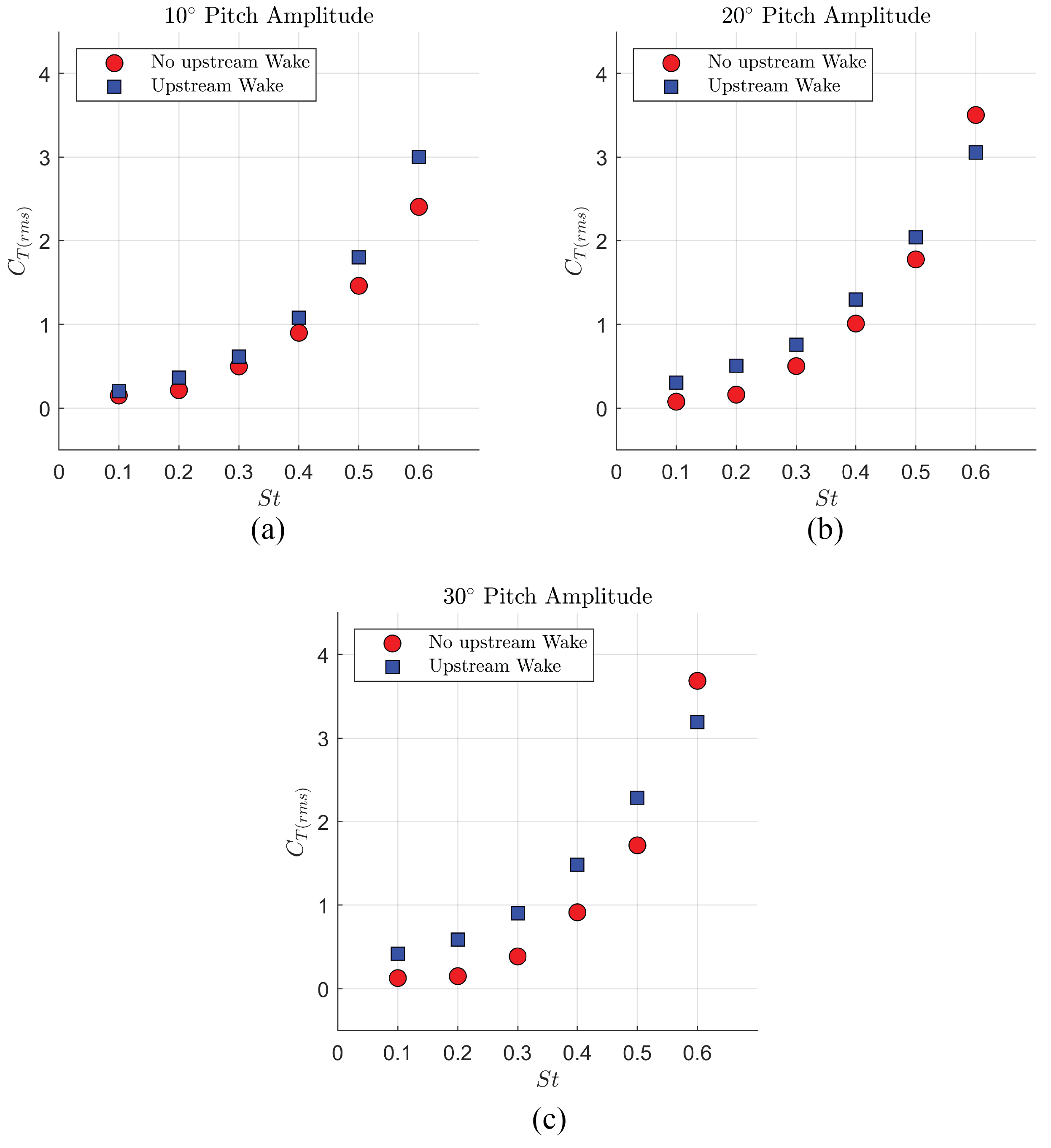}
    \caption{Root Mean Square (RMS) of Thrust coefficient, $C_T$, as a function of Strouhal number, $St$, for three different pitch amplitudes (10°, 20°, and 30°).}
    \label{fig:RMSThrust}
\end{figure}

\begin{figure}[hb!]
    \centering
    \includegraphics[width=1\linewidth]{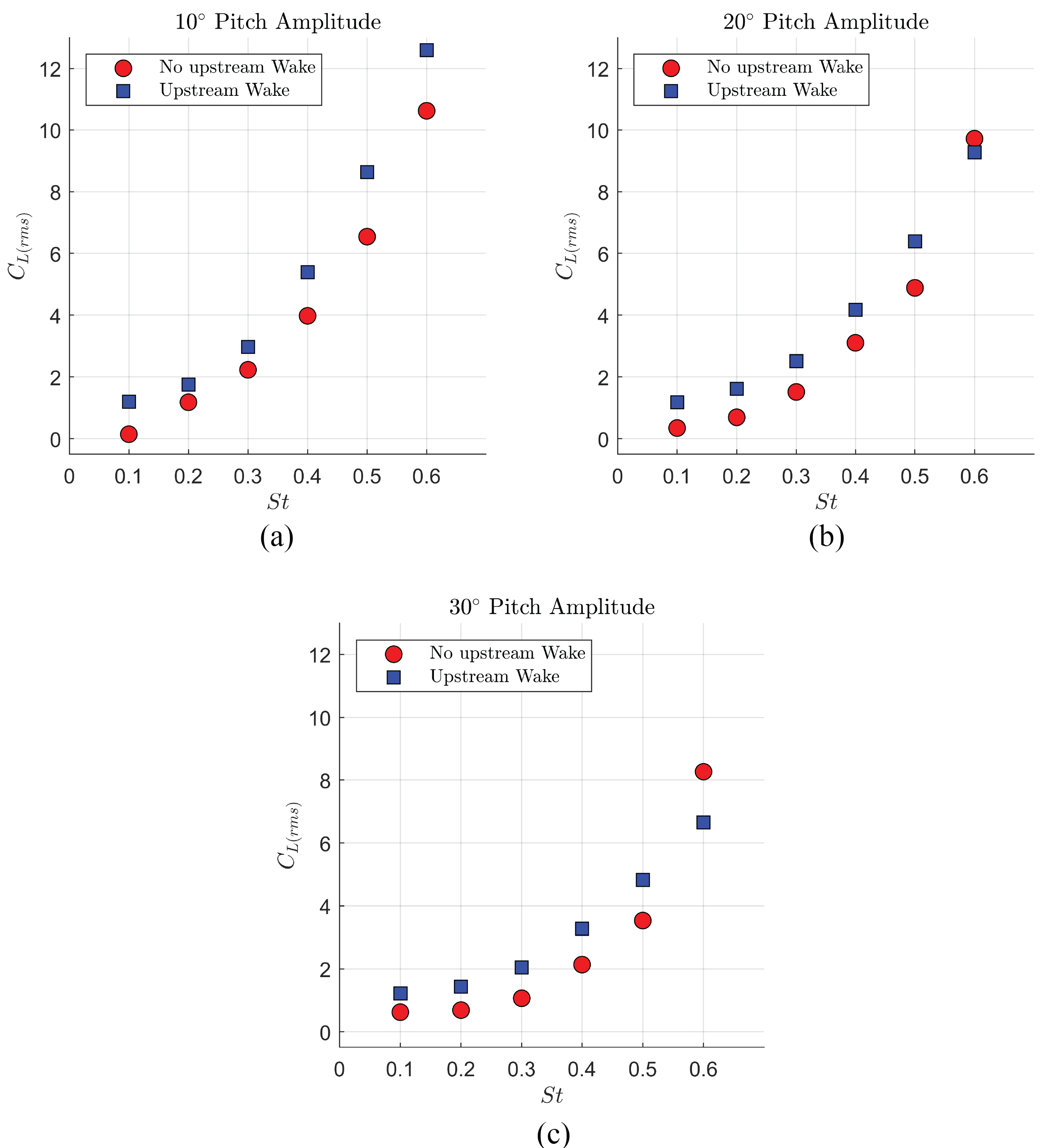}
    \caption{RMS for coefficient of Lift, $C_L$, side forces varying with different Strouhal number, $St$, for three different pitch amplitudes (10°, 20°, and 30°). }
    \label{fig:RMS_Lift}
\end{figure}

There is a direct correlation between the pitching amplitude and the flapping airfoil's performance. As the Strouhal number ($St$) grows, the findings reveal a continuous increase in thrust and side forces for decreasing pitching amplitudes, especially 10°as can be seen in Figure \ref{fig:RMSThrust} and Table \ref{tab:RMS_Thrust}. For instance, in both the no-wake and in-wake scenarios, thrust reaches its maximum levels as pitching amplitude increases. Airfoil side forces are quite sensitive to variations in pitching amplitude, according to the data. In comparison to the other pitching amplitudes, the side forces rise significantly at 10°, with percentage increases being the most notable as can be seen in Figure \ref{fig:RMS_Lift} and Table \ref{tab:RMS_Side_Forces_Airfoil}.\\

A more complex airfoil behavior is observed for pitching amplitudes of 20° and 30°. Thrust reaches a peak and subsequently decreases at $St = 0.60$; above 20°, there is great fluctuation, but the thrust coefficient still grows with $St$. At 30°, this variation becomes more noticeable, and the thrust exhibits substantial changes, indicating that flow dynamics are at work. Identical patterns are seen in side forces for pitching amplitudes of 20° and 30°, along with heightened variability and the possibility of a saturation effect, which becomes particularly apparent at $St = 0.60$. \\

The link between $St$ and pitching amplitude is generally increasing with $St$, although there are oscillations in the maximum thrust values at higher amplitudes (20° and 30°) that indicate that the relationship becomes less effective with bigger amplitudes. At 30°, the connection between maximum thrust and $St$ becomes less clear, which might be a result of the negative consequences of higher flow complexity and oscillation amplitudes.

\subsection{Performance Dependence on Strouhal Number}
An important factor determining the airfoil's performance is the Strouhal number. In the majority of instances, the creation of thrust and side force is enhanced when $St$ is increased from 0.10 to 0.60 across all pitching amplitudes as seen in Figure \ref{fig:RMSThrust}. Thrust is positively correlated with $St$ at a 10° pitching amplitude; higher $St$ values produce more thrust, but the rate of improvement slows down as the numbers get higher. $St$ improves thrust output, as this tendency persists under both no-wake and in-wake situations.\\

At greater pitching amplitudes (20° and 30°), the relationship between $St$ and airfoil performance gets increasingly complex as $St$ grows. At 20 degrees, the thrust grows with $St$ at first, but it peaks at around $St = 0.50$ and then drops at about $St = 0.60$. Side forces as shown in Figure \ref{fig:RMS_Lift}, exhibit a comparable pattern at this pitching amplitude, indicating that greater $St$ values may result in a saturation point or negative interactions. With side forces falling at $St = 0.60$ and more unpredictable behavior compared to lesser amplitudes, the trends remain complex at a 30° pitching amplitude. Increasing $St$ usually makes things work better, but it's possible that their impact will be less noticeable at bigger pitching amplitudes because of separation effects and possible flow instabilities at high oscillation frequencies.

\section{Conclusion and Future Recommendations}
This study offers essential insights into the performance dynamics of a flapping airfoil situated in the wake of an oscillating cylinder. The research reveals a notable enhancement in thrust generation and underscores the complex interplay between Strouhal number and pitching amplitude in influencing thrust efficacy. Although elevated Strouhal numbers typically augment thrust, their effect is closely related to the pitching amplitude. Reduced pitching amplitudes provide greater advantages from elevated Strouhal numbers, while increased amplitudes result in non-linear dynamics, potentially leading to diminishing returns.\\

The research indicates that increased pitching amplitude generally leads to greater side forces and thrust coefficients under wake situations. Lower Strouhal numbers indicate significant enhancements in both metrics; but, as the Strouhal number climbs, these advantages diminish, with certain instances even demonstrating declines. This pattern indicates that although heightened pitching amplitude can improve performance, there may be limits beyond which additional increases do not result in proportional enhancements.\\

Future research should be focus a detailed analysis of the influence of varying vortex shedding frequencies of the cylinder and other structural parameters on the interaction between the airfoil and the wake. These investigations may provide a deeper understanding of these relationships and lead to more advanced design methodologies, hence enhancing performance across many applications.

\begin{table}[hb!]
    \centering
    \caption{Root-Mean-Square of Thrust coefficient of the Airfoil at Different Strouhal numbers and pitching amplitudes}
    \begin{tabular}{|c|c|c|c|}
        \hline
        \textbf{St} & \textbf{No Wake} & \textbf{In Wake} & \textbf{Change} \\
        \hline
        \hline
        \multicolumn{4}{|c|}{\textbf{10$^\circ$ Pitching Amplitude}} \\ 
        \hline

        0.10 & 0.1506 & 0.2025 & 34.46\% \\
        0.20 & 0.2142 & 0.3646 & 70.21\% \\
        0.30 & 0.4969 & 0.6130 & 23.36\% \\
        0.40 & 0.8981 & 1.0789 & 20.13\% \\
        0.50 & 1.4594 & 1.8007 & 23.39\% \\
        0.60 & 2.4019 & 3.0025 & 25.01\% \\
        \hline
        \hline
        \multicolumn{4}{|c|}{\textbf{20$^\circ$ Pitching Amplitude}} \\ 
        \hline
        0.10 & 0.0765 & 0.3036 & 296.86\% \\
        0.20 & 0.1604 & 0.5052 & 214.96\% \\
        0.30 & 0.5012 & 0.7563 & 50.90\% \\
        0.40 & 1.0088 & 1.2953 & 28.40\% \\
        0.50 & 1.7746 & 2.0389 & 14.89\% \\
        0.60 & 3.5029 & 3.0581 & -12.70\% \\
        \hline
        \hline
        \multicolumn{4}{|c|}{\textbf{30$^\circ$ Pitching Amplitude}} \\ 
        \hline
        0.10 & 0.1274 & 0.4189 & 228.81\% \\
        0.20 & 0.1501 & 0.5874 & 291.34\% \\
        0.30 & 0.3859 & 0.8999 & 133.20\% \\
        0.40 & 0.9122 & 1.4848 & 62.77\% \\
        0.50 & 1.7169 & 2.2854 & 33.11\% \\
        0.60 & 3.6850 & 3.1914 & -13.39\% \\
        \hline
    \end{tabular}
    \label{tab:RMS_Thrust}
\end{table}

\begin{table}[hb!]
\centering
\caption{Root-Mean-Square of Side Forces of the Flapping Airfoil at Different Strouhal Numbers and Pitching Amplitudes}
\begin{tabular}{|c|c|c|c|}
\hline
\textbf{St} & \textbf{No Wake} & \textbf{In Wake} & \textbf{Change} \\
\hline
\hline
\multicolumn{4}{|c|}{\textbf{10$^\circ$ Pitching Amplitude}} \\ 
\hline
0.10 & 0.1447 & 1.1912 & 723.22\% \\
0.20 & 1.1786 & 1.7534 & 48.77\% \\
0.30 & 2.2301 & 2.9736 & 33.34\% \\
0.40 & 3.9779 & 5.3882 & 35.45\% \\
0.50 & 6.5433 & 8.6344 & 31.96\% \\
0.60 & 10.6192 & 12.5958 & 18.61\% \\
\hline
\hline
\multicolumn{4}{|c|}{\textbf{20$^\circ$ Pitching Amplitude}} \\ 
\hline
0.10 & 0.3435 & 1.1804 & 243.64\% \\
0.20 & 0.6913 & 1.6161 & 133.78\% \\
0.30 & 1.5121 & 2.5091 & 65.93\% \\
0.40 & 3.1036 & 4.1731 & 34.46\% \\
0.50 & 4.8834 & 6.3935 & 30.92\% \\
0.60 & 9.7144 & 9.2754 & -4.52\% \\
\hline
\hline
\multicolumn{4}{|c|}{\textbf{30$^\circ$ Pitching Amplitude}} \\ 
\hline
0.10 & 0.6246 & 1.2181 & 95.02\% \\
0.20 & 0.6905 & 1.4319 & 107.37\% \\
0.30 & 1.0670 & 2.0428 & 91.45\% \\
0.40 & 2.1347 & 3.2790 & 53.60\% \\
0.50 & 3.5358 & 4.8242 & 36.44\% \\
0.60 & 8.2692 & 6.6534 & -19.54\% \\
\hline
\end{tabular}
\label{tab:RMS_Side_Forces_Airfoil}
\end{table}

\bibliographystyle{IEEEtran}
\bibliography{conference.bib}

\end{document}